\documentclass[twocolumn,showpacs,preprintnumbers,amsmath,amssymb]{revtex4}

\usepackage{graphicx}
\usepackage{dcolumn}
\usepackage{bm}
\usepackage{epsfig}

\begin{document}

\title{Unusual criticality of $\mathrm{Cu_2Te_2O_5Br_2}$ under pressure}

\author{J. Kreitlow}
\email{j.kreitlow@tu-bs.de}
\author{S. S\"ullow}
\affiliation{
Institut f\"ur Metallphysik und Nukleare Festk\"orperphysik, TU Braunschweig, 38106 Braunschweig, Germany}

\author{D. Menzel}
\author{J. Schoenes}
\affiliation{Institut f\"ur Halbleiterphysik und Optik, TU Braunschweig, 38106 Braunschweig, Germany
}

\author{P. Lemmens}
\affiliation{Max-Planck-Institut f\"ur Festk\"orperforschung, 70569 Stuttgart, Germany
}

\author{M. Johnsson}
\affiliation{Department of Inorganic Chemistry, Stockholm University, S-10691 Stockholm, Sweden
}

\date{\today}

\begin{abstract}
We present measurements of the magnetic susceptibility $\chi (T)$ on
$\mathrm{Cu_2Te_2O_5Br_2}$ under externally applied pressure. From our data
we extract the pressure response of the antiferromagnetic phase transition
at $T_0=11.6$ K and of the overall magnetic coupling strength. Our
experiments indicate that with pressure the overall magnetic coupling
strength increases by about 25\% with applied pressure of only $\sim$8
kbar. In contrast, the phase transition temperature $T_0$ is significantly
suppressed and not observable anymore at a pressure of already 8.2 kbar.\end{abstract}

\pacs{75.10.Jm, 75.40.Cx, 75.40.Gb}
\maketitle

\begin{figure}[ht!]
\begin{center}
\epsfig{width=0.4\textwidth,file=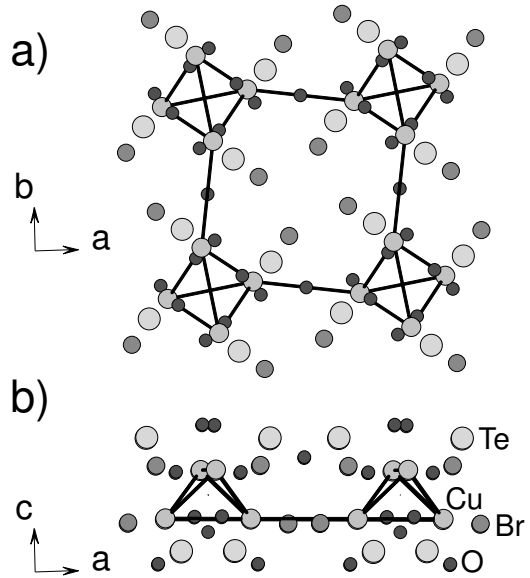}
\end{center}
\caption{\label{struc} A view of the crystal structure of $\mathrm{Cu_2Te_2O_5Br_2}$
onto the crystallographic $a$-$b$ (a) and $a$-$c$ plane (b) as
illustration for the planar arrangement of the Cu$^{2+}$ tetrahedra.}
\end{figure}

The spin-tetrahedra system $\mathrm{Cu_2Te_2O_5Br_2}$ \cite{john} belongs
to a class of quantum magnets which has been in the focus of intense
research efforts recently \cite{lemm,lemm2,pres,zaha,solo}. Here, the
presence of a spin gap through dimerization for a quantum magnet does not
lead to a non-magnetic singlet ground state. Instead, based on
thermodynamic and spectroscopic techniques an unusual magnetic ground state
has been evidenced. Tetragonal $\mathrm{Cu_2Te_2O_5Br_2}$ contains clusters
of Cu$^{2+}$ with $S$\,=\,1/2 in a distorted square planar CuO$_3$Br
coordination (Fig.~\ref{struc}). These tetrahedra form weakly coupled
sheets within the crystallographic $a$-$b$-plane. Therefore, this system is
ideal to study the interplay between the spin frustration of a tetrahedron
with localized low energy excitations and the tendency for a more
collective magnetism induced by inter-tetrahedra couplings.

The thermodynamic properties of $\mathrm{Cu_2Te_2O_5Br_2}$ are ascribed to
two magnetic couplings within the tetrahedra, with the competing exchange
constants $J_1$ and $J_2$ \cite{lemm}, and an inter-tetrahedra coupling
$J_c$\cite{gros}. As result of the coupling, the system undergoes a phase
transition at $T_0=11.6$ K. Neutron powder diffraction of
$\mathrm{Cu_2Te_2O_5Br_2}$ reveals an antiferromagnetically ordered state
with a strongly reduced magnetic moment of 0.51(5)$\mu_B/Cu^{2+}$ below
$T_0$ \cite{zaha}. On a microscopic level the cause for the phase
transition has not unambiguously been resolved
\cite{bren,vale,gros,jens,tots,koto}. In particular, with the existence of
low lying excitations a magnetically ordered state close to quantum
criticality has been discussed.

\begin{figure}[ht!]
\begin{center}
\epsfig{width=0.5\textwidth,file=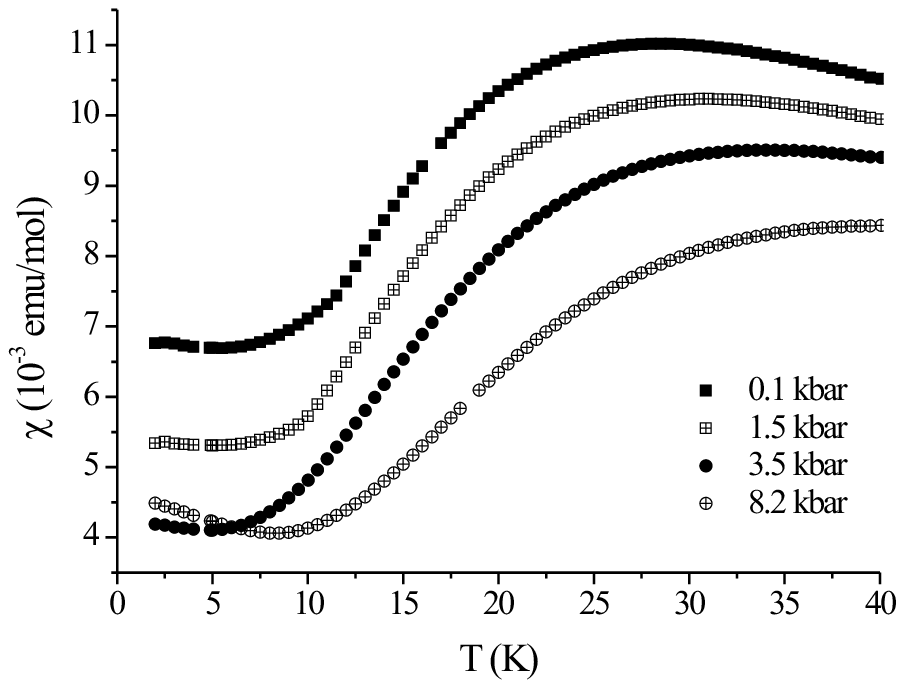}
\end{center}
\caption{\label{chi} The magnetic susceptibility $\chi (T)$ of
$\mathrm{Cu_2Te_2O_5Br_2}$ as function of pressure, measured in an
external field of 5 T.}
\end{figure}

Pressure experiments have proven to be a particularly useful tool
to study quantum critical behavior. Therefore, in this work we
present a pressure study on $\mathrm{Cu_2Te_2O_5Br_2}$. For our
experiments we used a CuBe pressure cell in a commercial SQUID
magnetometer to measure $\chi (T)$ at pressures up to 8.2 kbar and
in external fields up to 5 T for temperatures 2 - 40 K. A powder
sample, which has been prepared as described in Ref. \cite{john},
was pressed together with GE-Varnish into a pellet, which was
placed in the middle of a teflon tube. The tube was filled with a
hydraulic pressure medium (FC-77) and was loaded into the CuBe
pressure cell. In Fig.~\ref{chi} we plot a set of representative
measurements on $\mathrm{Cu_2Te_2O_5Br_2}$ with our cell for
pressures up to 8.2 kbar.

According to ambient pressure experiments \cite{lemm}, the
magnetic susceptibility $\chi (T)$  exhibits a broad maximum at
about $T_{Max}=30$ K. This susceptibility maximum represents a
measure for the overall magnetic coupling strength, {\it viz.}, the
size of $J_1$ and $J_2$. Below $T_{Max}$ a strong reduction of
$\chi$ occurs, as is typical for the onset of antiferromagnetic
correlations. The ordering temperature $T_0$ is identified as a
step in the temperature derivative $\partial \chi/\partial T$.

\begin{figure}[ht!]
\begin{center}
\epsfig{width=0.5\textwidth,file=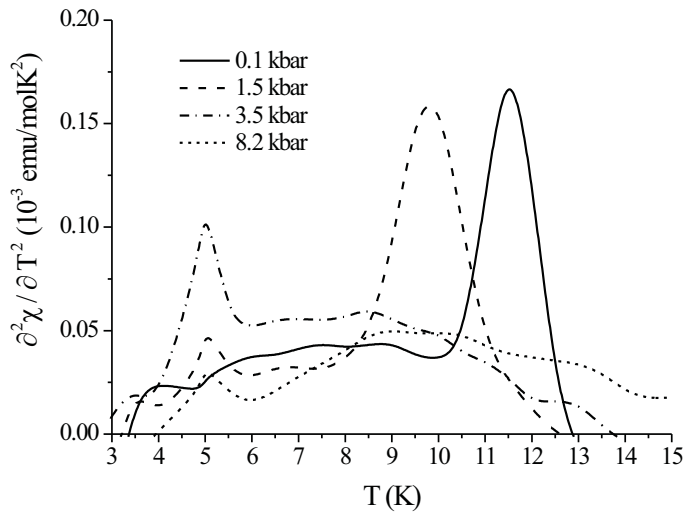}
\end{center}
\caption{\label{abl} The second derivative $\partial
^2\chi/\partial T^2$ of $\mathrm{Cu_2Te_2O_5Br_2}$ as function of
pressure in an applied field of 5 T.}
\end{figure}

Our data at 0.1 kbar closely resemble the ambient pressure
behavior from Ref.\cite{lemm}. Since a step in $\partial
\chi/\partial T$ corresponds to a maximum in $\partial
^2\chi/\partial T^2$, we determine $T_0$ at 0.1 kbar from the
latter quantity to 11.6 K, in good agreement with Ref.\cite{lemm}
(Fig.\ref{abl}).

With increasing pressure the phase transition temperature $T_0$ decreases,
and we obtain $T_0=9.8 $K at 1.5 kbar (Fig.\ref{abl}). At 3.5 kbar the
maximum in $\partial ^2\chi/\partial T^2$ has shifted to 5 K. However,
since for both the data taken at 1.5 kbar and 8.2 kbar a similar, but much
smaller maximum in $\partial ^2\chi/\partial T^2$ appears, this feature at
5 K might possibly be induced by a changing cooling mode of the SQUID in
this temperature range.

Alternatively, it could be argued that for the 3.5 kbar measurement the
broad anomaly underlying the peak at 5 K represents a remnant of the
antiferromagnetic ordering. In that case the data would indicate a range of
$T_0$ between 5 and 8 K. Still, the observation of similar broad anomalies
for the higher pressure experiments seems to speak against such
interpretation.

\begin{figure}[ht!]
\begin{center}
\epsfig{width=0.5\textwidth,file=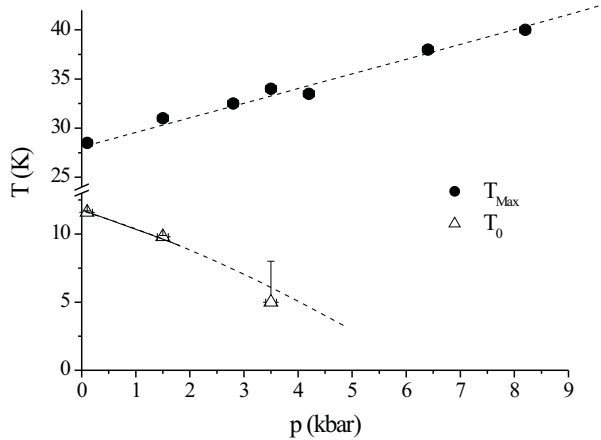}
\end{center}
\caption{\label{max} The pressure dependence of $T_{Max}$ and
$T_0$ in an external field of 5 T. Lines are guides to the eye.}
\end{figure}

In Fig.\ref{max} we summarize the pressure dependence of $T_0$. The error
bar at 3.5 kbar reflects the uncertainty about the determination of $T_0$
at this pressure. Altogether, the data suggest a suppression of $T_0$ in
the range 5-8 kbar. This statement is supported by the absence of any clear
signature of magnetic ordering for the measurement at 8.2 kbar. Hence, our
experiments indicate that $\mathrm{Cu_2Te_2O_5Br_2}$ is situated in the
proximity to a nonmagnetic phase. As the constituting unit is a tetrahedron
with antiferromagnetic exchange interaction this phase is suggested to be
identical with a short range correlated singlet phase. However also other
scenarios have been put forward based on theoretical arguments \cite{tots}.
Two experimental finding shine further light on the peculiarity of
$\mathrm{Cu_2Te_2O_5Br_2}$. First is the observation of a related
instability with a much higher $T_0$=18.2~K in $\mathrm{Cu_2Te_2O_5Cl_2}$
that has a 7\% smaller unit cell volume \cite{lemm}. This compound also has
a completely different low energy excitation spectrum in light scattering
experiments \cite{lemm2}. The second observation is the evidence for an
incommensurate ordering vector of $\mathrm{Cu_2Te_2O_5Br_2}$ for T$<$$T_0$
\cite{zaha}. Both experimental results imply that the ordering temperature
is not only given by a mean field-like inter-tetrahedra coupling and that
some additional effect, most probably some antisymmetric
Dzyaloshinskii-Moriya (DM) interaction plays some role to establish long
range ordering.

Moreover, from $\chi (T)$ we derive the pressure dependence of the
susceptibility maximum. It increases from $T_{Max}=28.5$ K (determined via
$\partial \chi/\partial T$ = 0 ) at 0.1 kbar applied pressure to
$T_{Max}=40$ K at 8.2 kbar (Fig.\ref{max}). This increase indicates a very
substantial strengthening of the intra-tetrahedra coupling with applied
pressure, yielding an enhancement of 25\% at highest applied pressure.

The contrasting pressure response of $T_0$ and $T_{Max}$ is very unusual
and likely reflects competing energy scales. If the ordering temperature
$T_0$ would be only controlled by the overall magnetic coupling strength,
we would expect an increase of $T_0$ with $T_{Max}$. Therefore, the
decreasing $T_0$ possibly is the result of enhanced frustration $J_1/J_2$
on the tetrahedra. Another scenario would be a weakened inter-tetrahedra
coupling $J_c$ with pressure. This however seems unlikely as under pressure
the inter-tetrahedra distance decreases which in the absence of structural
symmetry modifications should lead to an increase of $J_c$.

In summary, we have performed a pressure study on the susceptibility of
$\mathrm{Cu_2Te_2O_5Br_2}$. We have determined the pressure response of the
antiferromagnetic phase transition temperature $T_0$ and the overall
magnetic coupling strength. While we find a strengthening of the magnetic
coupling with pressure, attributed to intra-tetrahedra exchange pathes,
antiferromagnetic order is rapidly suppressed. Tentatively, we relate this
behavior to an enhancement of frustration or a weakening of antisymmetric
interactions in the system. However, to weight such scenarios
$\mathrm{Cu_2Te_2O_5Br_2}$ additional thermodynamic and spectroscopic
pressure experiments to higher pressure, structural studies under pressure
and following theoretical investigations will be necessary. Such work is in
preparation.

This work was supported by the Deutsche Forschungsgemeinschaft DFG under
projekt number SU/6-1, SPP1073, and INTAS 01-278.

\end{document}